\documentclass[aps,prb,twocolumn,groupedaddress,showpacs,superscriptaddress,amssymb,amsmath,noeprint,longbibliography]{revtex4-2}
\usepackage{graphicx}
\usepackage{dcolumn}
\usepackage{bm}
\usepackage{hyperref}
\usepackage{cleveref}
\hypersetup{
    colorlinks=true,
    linkcolor=blue,
    urlcolor=blue,
	citecolor=blue
}
\usepackage{comment}
\usepackage{color}
\usepackage[utf8]{inputenc}
\usepackage{graphicx}
\usepackage{tabularx}
\usepackage{xcolor}
\usepackage{amsmath}
\usepackage{dcolumn}
\usepackage{bm}
\usepackage{epsf}
\usepackage{braket}
\usepackage{tensor}
\usepackage{soul}
\usepackage{bbm}
\usepackage{tikz}
\usepackage[normalem]{ulem}

\begin{document}

\title{Quantum Mpemba effect for operators in open systems}

\author{Pitambar Bagui}
\email{pitambar.bagui@students.iiserpune.ac.in} 
\affiliation{Department of Physics, Indian Institute of Science Education and Research, Pune 411008, India}

\author{Bijay Kumar Agarwalla}
\email{bijay@iiserpune.ac.in}
\affiliation{Department of Physics, Indian Institute of Science Education and Research, Pune 411008, India}

\date{\today} 

\begin{abstract}
The quantum Mpemba effect concerns with anomalous relaxation of quantum states that evolves either under unitary or non-unitary dynamics. In the context of open quantum systems, while most studies focus on quantum states evolving under completely positive trace-presing dynamics described by the Gorini–Kossakowski–Sudarshan–Lindblad (GKSL) master equation, we demonstrate that an analogous effect can arise at the level of operators. In particular, we show that operators that evolves under the adjoint Liouvillian -- despite not being a trace-preserving map -- can still exhibit a genuine Mpemba effect. We derive general conditions under which this phenomenon can occur and validate our predictions for three different open quantum setups. Our results broaden the scope of the Mpemba effect in quantum systems and provide a framework for controlling the relaxation of physically relevant observables.
\end{abstract}

\maketitle

{\textit{Introduction.--} 
Understanding the relaxation dynamics of quantum systems has long been a subject of fundamental research interest \cite{Teza_prl_2026,Demler_science_2012,Izrailev_prl_2012,Marko_pre_2015,Solomon_pra_2004,Vladimir_pra_2018}. The way a quantum system relaxes to its asymptotic steady-state, not only provides insight into the underlying physical mechanisms, but also plays a pivotal role in shaping the emerging field of quantum technology \cite{Xin_prl_2026,Jen_2026_arxiv,Bruss_pra_2006,Gerhard_pra_2003,rensen_prl_2011}. Despite significant progress in this direction, the study of non-equilibrium quantum dynamics continues to be an active area of research, particularly in the context of open quantum systems \cite{breuer2002theory,lidar2020} where the system of interest continuously interacts with the surrounding.

Recent theoretical studies in both isolated \cite{Ares_calarese_2023_Nature,Rylands_Bruno_2024_PRL,Shion_calabrese_2024_PRB,Murciano_calabrese_2024_JSM,Lastres_calabrese_JSM_2025,Ares_Sara_2025_PRB,Yamashika_filiberto_2025_PRA,Chalas_calabrese_2024_JSM,Zhang_2026_entropy,Jian_PRB_2026,Heng_2025_PRB,Bhore_clerk_2025,hallam_2025} and open quantum systems \cite{PhysRevLett.131.080402,bagui2025detection,bao2025,PhysRevA.106.012207,PhysRevResearch.6.033330,PhysRevLett.133.136302,PhysRevLett.131.080402,longhi2026pontus,PhysRevLett.134.220403,ganguly2026,PhysRevLett.133.140404,longhi2024bosonic,pan_2026,zhang_chen_2026,Longhi_Pontous_skin_2026,Wei2026,Aditi_sende_2026,chattopadhyay_2026,solanki_2025,Ares_review_2025,chatterjee_2025_direct,Zhang2025_observation_Mpemba,PhysRevLett.133.010403,Zala_2025,alishahiha_2025} have shown that a system initially farther from equilibrium can relax faster than one that is closer to it. This counterintuitive behavior is the quantum analogue of the Mpemba effect, originally brought to attention by Erasto B. Mpemba, who observed that hot water can freeze faster than cold \cite{E_B_Mpemba_1969}. Since then, the effect has been extensively studied in classical systems \cite{Greaney2011,PhysRevLett.119.148001,Hu2018,Schwarzendahl_2022,PhysRevLett.124.060602,bera_2023_vucleja,Vynnycky_2010}, with various mechanisms proposed to explain its origin.

In the quantum regime, the Mpemba effect has attracted growing interest, with studies demonstrating its emergence in both isolated \cite{Ares_calarese_2023_Nature,Rylands_Bruno_2024_PRL,Shion_calabrese_2024_PRB,Murciano_calabrese_2024_JSM} and open systems \cite{PhysRevLett.131.080402,bagui2025detection,bao2025}. In isolated systems, it emerges in the context of symmetry considerations: for two initial pure states that break a conserved $U(1)$ symmetry of an integrable Hamiltonian to different extents, the state with higher symmetry breaking can exhibit faster symmetry restoration at the subsystem level \cite{Ares_calarese_2023_Nature}. On the other hand, in open systems, the effect is controlled by the spectrum of the dynamical generator and the overlap with slowly decaying modes \cite{PhysRevLett.131.080402}. An initial state farther from the steady state, but orthogonal to the slowest decay mode (SDM), can relax faster than a closer state with finite overlap with the SDM \cite{PhysRevLett.131.080402}.

Existing studies so-far have largely focused on the relaxation of quantum states \cite{Ares_calarese_2023_Nature,PhysRevLett.131.080402}. However, often experiments probe the relaxation of specific physical properties \cite{PhysRevX.7.031027,PhysRevE.90.012121,Alpino_2025}, represented by self-adjoint operators, rather than the full quantum state. This naturally raises the question of whether the Mpemba effect can manifest at the level of observables. Recent work \cite{bagui_2025} has taken initial steps in this direction by formulating the problem in operator space. However, a key challenge remains: the steady state of the operator dynamics generally depends on the choice of the initial operator \cite{bagui_2025}. This dependence complicates a direct extension of the Mpemba effect from states to observables and motivates the need for a framework in which the notion of steady-state relaxation is independent of the initial operator.

In this work, we develop a framework to characterize the Mpemba effect directly at the level of observables. By formulating the dynamics in operator space, we identify suitable operations under which physical observables can exhibit accelerated relaxation analogous to the traditional quantum state Mpemba effect. In particular, we address a key limitation of previous approaches, namely the dependence of the steady state on the initial operator \cite{bagui_2025}, and provide a formulation in which both the operator and the transformed operator relax to the same steady state. Within this framework, we show that the relaxation speed of an observable is governed by its overlap with its slowest decay modes, and that appropriate transformations can bypass the overlap \cite{PhysRevLett.131.080402}, leading to an exponential speed-up in the relaxation and faster convergence towards the steady state value. We further provide conditions on distance measure to observe Mpemba effect for operators. Our results establish that the Mpemba effect is not restricted to quantum states, but can manifest in experimentally relevant observables as well.

{\textit{Recipe for observing genuine Mpemba effect for operators.--}}
The time evolution of a quantum state interacting with the external environment is described by a quantum master equation derived under the standard Born-Markov and secular approximations \cite{breuer2002theory,lidar2020,schaller2014open}. Under these approximations, the reduced dynamics of the system is governed by the  Gorini-Kossakowski-Sudarshan-Lindblad (GKSL) master equation \cite{breuer2002theory}, which provides the most general form of a Markovian, completely positive and trace-preserving quantum evolution. The GKSL equation is given by
\begin{align}
\!\!\frac{d\rho}{dt}\!=\!\mathcal{L}\rho \equiv -i[H,\rho]\!+\!\sum_k\gamma_k\big(L_k\rho L_k^{\dagger}-\frac{1}{2}\{L_k^{\dagger}L_k,\rho\}\big),
    \label{lindblad_state}
\end{align}
where the first term provides the unitary contribution with $H$ being the Hamiltonian of the system. The second term represents the dissipative contribution with $L_k$ being the jump operator for the $k$-th channel and $\gamma_k$ is the associated rate of dissipation. Since the Liouvillian $\mathcal{L}$ is a non-Hermitian superoperator, its spectral property is characterized by the left and right eigenoperators. These are defined through 
$
\mathcal{L}r_k=\lambda_kr_k,~~~\mathcal{L}^{\dagger}l_k=\lambda_k^{*}l_k$ with $\lambda_k$ being the eigenvalue of the Liouvillian. The left and right eigen-operators together form the complete basis and obey the bi-orthonormality condition, i.e., $\mathrm{Tr}(l_ir_j)=\delta_{ij}$ \cite{PhysRevA.98.042118,lidar2020,breuer2002theory}.

In comparison to the state, the operator evolves in the Heisenberg picture, under the adjoint-Liouvillian $\mathcal{L}^{\dagger}$ \cite{breuer2002theory}. The time evolution of an operator $\mathcal{O}$ is given by $\mathcal{O}(t)=e^{\mathcal{L}^{\dagger}t}\mathcal{O}(0)$, where $\mathcal{L}^{\dagger}$ reads as,
\begin{equation}
\mathcal{L}^{\dagger}\mathcal{O}=i[H,\mathcal{O}]+\sum_{k}\gamma_k \big(L_k^{\dagger}\mathcal{O}L_k-\frac{1}{2}\{L_k^{\dagger}L_k,\mathcal{O}\}\big).
    \label{operator_evolution}
\end{equation}
Following the spectral decomposition of $\mathcal{L}^{\dagger}$, the time evolved operator $\mathcal{O}(t)=e^{\mathcal{L}^{\dagger}t}$ can be expressed as,
\begin{equation}
    \mathcal{O}(t)=\mathcal{O}_{\mathrm{ss}}+\sum_i \mathrm{Tr}\big[r_i\mathcal{O}(0)\big]e^{\lambda_i t}l_i,
\end{equation}
where $\mathcal{O}_{\mathrm{ss}}=\mathrm{Tr}\big[\rho_{\mathrm{ss}}\mathcal{O}(0)\big]$ and $\mathrm{Tr}\big[r_i\mathcal{O}(0)\big]$ is the overlap of the initial operator $\mathcal{O}(0)$ with the right eigenvector $r_i$. The time an operator takes to reach its steady state at the long-time scale is decided by its spectral gap $|\mathrm{Re}(\lambda_1)|$ and the overlap $\mathrm{Tr}(r_1\mathcal{O})$. Bypassing this mode will cause an exponential speed-up in the relaxation process of the operator $\mathcal{O}(t)$, and the dynamics at the long time scale will be governed through the real part of the next SDM. In contrast to the states, bypassing the slowest relaxation mode $r_1$ through a unitary transformation on the operators $\mathcal{O}(0)$, i.e., $\mathrm{Tr}\big[r_1 U^{\dagger}\mathcal{O}(0)U\big]=0$ also changes the asymptotic operator \cite{breuer2002theory,bagui_2025}. In other words, the steady-state value $\mathcal{O}_{\mathrm{ss}}=\mathrm{Tr}\big[\rho_{\mathrm{ss}}\mathcal{O}(0)\big]$ changes to $\widetilde{\mathcal{O}}_{\mathrm{ss}}=\mathrm{Tr}\big[\rho_{\mathrm{ss}}\widetilde{O}(0)\big]$ where  $\widetilde{O}(0)=U^{\dagger}\mathcal{O}(0)U$. Therefore, unlike for the quantum states, a unitarily transformed operator can relax faster but approaches to a different steady-state value and hence complicates the extension of the Mpemba effect from states to observables. Therefore, in order to achieve the same steady-state i.e., $\mathcal{O}_{\mathrm{ss}}= \widetilde{\mathcal{O}}_{\mathrm{ss}}$, one needs to go beyond the unitary transformation protocol. We here adopt a different approach that removes the contribution of the SDM from the operator while ensuring that the steady state remains unchanged. 
Specifically, we define the transformed operator as
\begin{equation}
\widetilde{\mathcal{O}}=\mathcal{O}-\mathrm{Tr}\big[r_1\mathcal{O}\big]l_1,
\label{mode_bypassing}
\end{equation}
where $r_1$ ($l_1$) denotes the right (left) eigenvector corresponding to the SDM of the operator. It is important to note that the SDM for an operator need not always be the same as the SDM of the Liouvillian. An important consequence of this construction is that the locality properties of the transformed operator $\widetilde{\mathcal{O}}$ depend on the structure of the removed SDM. In particular, if the corresponding left eigenoperator $l_1$ is local, then a local initial operator $\mathcal{O}$ remains local after the transformation. However, when the SDM is intrinsically nonlocal, the subtraction term $\mathrm{Tr}[r_1\mathcal{O}] \, l_1$ introduces nonlocal contribuition into $\widetilde{\mathcal{O}}$. Consequently, even if the original operator $\mathcal{O}$ is strictly local, the transformed operator can become highly nonlocal [for details see \cite{supp}]. 
Interestingly, the construction in Eq.~\eqref{mode_bypassing} preserves the steady state while enabling faster relaxation. In particular,
\begin{equation}
\widetilde{\mathcal{O}}_{\mathrm{ss}}
=\mathrm{Tr}\big[\widetilde{\mathcal{O}}\rho_{\mathrm{ss}}\big]\mathbb{I}
=\mathrm{Tr}\big[{\mathcal{O}}\rho_{\mathrm{ss}}\big]\mathbb{I}
-\mathrm{Tr}\big[r_1\mathcal{O}\big]\mathrm{Tr}\big[l_1\rho_{\mathrm{ss}}\big]\mathbb{I}.
\end{equation}
Due to the bi-orthonormality condition, $\mathrm{Tr}[l_1\rho_{\mathrm{ss}}]=0$ as $\rho_{\mathrm{ss}}=r_0$ and as a consequence  $\widetilde{\mathcal{O}}_{\mathrm{ss}}={\mathcal{O}}_{\mathrm{ss}}$. Therefore, the steady-state value of the operator before and after transformation remains unchanged. More generally, this procedure can be extended to eliminate multiple consecutive SDMs. In that case we can write the transformed operator as
\begin{equation}
\widetilde{\mathcal{O}}=\mathcal{O}-\sum_i \mathrm{Tr}\big[r_i\mathcal{O}\big]\,l_i,
\end{equation}
which leads to a further acceleration of the relaxation dynamics. Note that this procedure only guarantees faster convergence of the time-evolved operator $\mathcal{O}(t)$ to its asymptotic value. However, in order to probe the emergence of the Mpemba effect, it is essential to compare the initial distances: specifically, how far the transformed operator $\widetilde{\mathcal{O}}(0)$ lies from its steady-state $\mathcal{O}_{\rm ss}$ relative to the original operator $\mathcal{O}(0)$. This is what we investigate next.

\vspace{0.3em}
\textit{Distance measure and condition for Mpemba effect for operators.--}
The dynamical map $e^{\mathcal{L}t}$ governing the evolution of a quantum state is completely positive and trace preserving (CPTP) \cite{breuer2002theory}. As a consequence, contractive distance measures such as the trace distance and quantum relative entropy decay monotonically with time \cite{breuer2002theory,nielsen2010quantum,Ares_2025}. These measures are therefore commonly used to quantify the distance between the time-evolved state $\rho_t$ and the steady state $\rho_{\mathrm{ss}}$, and serve as standard tools to identify the standard Mpemba effect \cite{Ares_2025}. In contrast, operators evolve under the adjoint map $e^{\mathcal{L}^\dagger t}$, which is, in general, not trace preserving \cite{breuer2002theory} (except in the special case of unital dynamics)
As a result, standard distance measures such as the trace distance \cite{Ares_2025,nielsen2010quantum} do not necessarily exhibit monotonic decay under adjoint evolution, and hence are not suitable for analyzing the Mpemba effect at the level of operators.

Recent study ~\cite{bagui_2025} has shown that the \emph{dressed distance}, which can be interpreted as a trace distance under the Petz recovery map \cite{Liang_petz_map_2025,MarkM_petzmap_2025}, remains contractive even under a non-trace-preserving dynamics, generated by $e^{\mathcal{L}^\dagger t}$. This makes it a standard distance measure for studying the Mpemba effect in the operator framework. The dressed distance is defined as \cite{bagui_2025}
\begin{equation}
\mathcal{D}_{\mathrm{dd}}(\mathcal{O}, \mathcal{O}_{\mathrm{ss}})
= \mathrm{Tr}\left( \sqrt{X^\dagger X} \right),
\end{equation}
where $X = \sqrt{\rho_{\mathrm{ss}}}\,(\mathcal{O} - \mathcal{O}_{\mathrm{ss}})\,\sqrt{\rho_{\mathrm{ss}}}.
$
For unital dynamics ($\rho_{\mathrm{ss}} = \mathbb{I}/d$ with $d$ being the dimension of the Hilbert space), the dressed distance reduces to the standard trace distance up to a constant prefactor.

Following the mode-bypassing procedure introduced in Eq.~(\ref{mode_bypassing}), we construct a transformed operator $\widetilde{\mathcal{O}}$ by removing the contribution of its SDM. By construction, $\widetilde{\mathcal{O}}$ relaxes faster towards the steady state. The key question is whether this transformation also increases the initial distance from the steady-state, i.e.,
$
\mathcal{D}_{\mathrm{dd}}(\widetilde{\mathcal{O}}, \mathcal{O}_{\mathrm{ss}})
>
\mathcal{D}_{\mathrm{dd}}(\mathcal{O}, \mathcal{O}_{\mathrm{ss}}).
$
If this condition is satisfied at $t=0$, the operator can exhibit the Mpemba effect during the relaxation process. 
To analyze this further, we define
\begin{equation}
A = \sqrt{\rho_{\mathrm{ss}}}\,(\mathcal{O} - \mathcal{O}_{\mathrm{ss}})\sqrt{\rho_{\mathrm{ss}}}, 
\quad
B = \sqrt{\rho_{\mathrm{ss}}} \Big[  \mathrm{Tr}\big(r_1 \mathcal{O}\big) l_1 \Big] \sqrt{\rho_{\mathrm{ss}}},
\label{A&B}
\end{equation}
so that the change in the dressed distance can be expressed as
$
\Delta = \|A - B\| - \|A\|,
$
where $\|X\| = \mathrm{Tr}(\sqrt{X^\dagger X})$ is the trace norm.
Invoking the polar decomposition form for operators \cite{polar_decom,polar_Nicholas} $X = U_X |X|$ where $|X|= \sqrt{X^\dagger X}$, one can show that [for details see \cite{supp}]
\begin{equation}
\|X\| =  \mathrm{Tr}(U_X^\dagger X)\geq \mathrm{Re} \big[\mathrm{Tr}(U^{\dagger}X)\big],
\label{bound}
\end{equation}
where $U$ is an arbitrary unitary operator and the equality only holds for $U=U_X$. Applying the polar decomposition form and the inequality in Eq.~\eqref{bound} to $A$ and $B$ in Eq.~\eqref{A&B}, we obtain the lower bound on $\Delta$ as
\begin{align}
\label{lower-bound}
\|A-B\|&=\mathrm{Re}\big[\mathrm{Tr}[U^{\dagger}_{A-B}(A-B)]\big]\nonumber \\&\geq \mathrm{Re} \big[\mathrm{Tr}[U^{\dagger}_A(A-B)]\big] =\|A\| - \mathrm{Re} \big[\mathrm{Tr}[U^{\dagger}_A(B)]\big]\nonumber \\
\Delta &\geq - \mathrm{Re} \big[\mathrm{Tr}[U^{\dagger}_A(B)]\big],
\end{align}
where recall that $
\Delta = \|A - B\| - \|A\|.
$
Similarly, one can obtain an upper bound on $\Delta$ as
\begin{align}
\label{upper-bound}
  \|A\| &\geq  \mathrm{Re} \big(\mathrm{Tr}[U^{\dagger}_{A-B}(A)]\big)\nonumber \\&= \mathrm{Re} \big(\mathrm{Tr}[U^{\dagger}_{A-B}(A-B+B)]\big)\nonumber \\
   &\geq  \|A-B\| + \mathrm{Re} \big(\mathrm{Tr}[U^{\dagger}_{A-B} B]\big)\nonumber \\
  \Delta &\leq  - \mathrm{Re} \big(\mathrm{Tr}
  [U^{\dagger}_{A-B} B]\big).
\end{align}
From the above two conditions, as given in Eq.~\eqref{lower-bound} and Eq.~\eqref{upper-bound}, we receive the following inequality for $\Delta$
\begin{equation}
\label{inequality}
-\mathrm{Re}\Big[\mathrm{Tr}(U_A^\dagger B)\Big] \leq 
\Delta
\leq
-\mathrm{Re}\Big[\mathrm{Tr}(U_{A-B}^\dagger B)\Big].
\end{equation}
Therefore, a \textit{sufficient} condition for the increase of the dressed distance i.e.,  $\Delta \geq0$ is
$
\mathrm{Re}\,\big[\mathrm{Tr}(U_A^\dagger B)\big] \leq 0,
$
while a \textit{necessary} condition is given by 
$
\mathrm{Re}\,\big[ \mathrm{Tr}(U_{A-B}^{\dagger} B)\big] \leq 0.
$
Interestingly, the sufficient condition admits a simple geometric interpretation. The unitary operator $U_A$, obtained from the polar decomposition $A = U_A |A|$, captures the orientation of the operator $A$ in operator space. This is analogous to the phase factor in the polar representation of a complex number, $z = r e^{i\theta}$, where $e^{i\theta}$ determines the orientation of $z$ in the complex plane. The quantity $\mathrm{Tr}(U_A^\dagger B)
$ can be interpreted as the Hilbert-Schmidt inner product \cite{PhysRevLett.133.140404} $\langle U_A,B\rangle$  that measures the alignment between $A$ and the contribution $B$ arising from the removed mode. When this quantity is negative, it indicates that $B$ is oppositely oriented to $A$ in operator space. Consequently, subtracting such a contribution (i.e., bypassing these modes) increases the norm $\|A - B\|$ relative to $\|A\|$, leading to an increase in the dressed distance. This provides an intuitive picture: removing components that are anti-aligned with the operator enhances its initial distance from the steady state, thereby enabling the Mpemba effect for operators. In what follows, we illustrate the genuine Mpemba effect for operators for three paradigmatic open quantum setups.

\vspace{0.3em}


\textit{Example 1: Single-qubit evolving under a unital map.--}
As a first example, we consider a single qubit with Hamiltonian
$ H=\omega_0 \sigma_z/2$,
subjected to dissipation through the jump operator $\sigma_x$ with dissipation strength $\gamma$. In this case, the steady state is maximally mixed, $\rho_{\mathrm{ss}}=\mathbb{I}_2/2,
$
and the corresponding dynamical map is unital. The Liouvillian eigenvalues are given by
$0, -\gamma \pm \sqrt{\gamma^2-\omega_0^2}$, and $-2\gamma
$. Depending on the relative magnitudes of $\gamma$ and $\omega_0$, the spectrum can consist of purely real eigenvalues, a complex-conjugate pair, or an exceptional point of second order. Here, we focus on the regime $\gamma>\omega_0$, where all eigenvalues are real.
\begin{figure}
    \centering
    \includegraphics[width=\columnwidth]{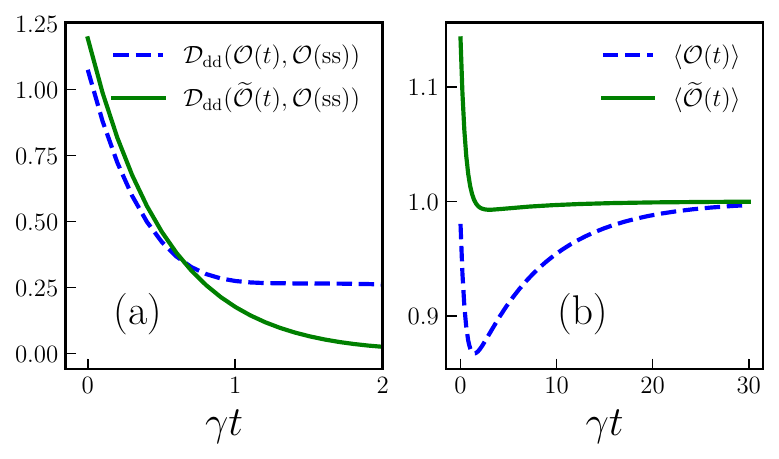}
    \caption{Presence of the genuine operator Mpemba effect in a single qubit setup evolving under a unital map. (a) Time evolution of the dressed distance for the original operator $\mathcal{O}=b_0\mathbb{I}+\vec{b}\cdot\vec{\sigma}$ and the transformed operator $\widetilde{\mathcal{O}}$, obtained by bypassing the SDM. Although $\widetilde{\mathcal{O}}$ is initially farther from the asymptotic value, it relaxes significantly faster due to the removal of the SDM. (b) Time evolution of the corresponding expectation values evaluated in the initial state $\ket{\psi_0}=\ket{+}$, demonstrating the accelerated relaxation of the transformed operator $\widetilde{\mathcal{O}}$. The parameters used are $\omega_0=0.5$, $\gamma=1$, $b_0=1$, and the vector $\vec{b}$ is chosen randomly with values for the components given as $(-0.01954,\, 0.52816,\,0.09795)$.}
\label{Mpemba_unital}
\end{figure}
In this regime, the SDM of the Liouvillian $\mathcal{L}$ corresponds to the eigenvalue
$
\lambda_1=-\gamma+\sqrt{\gamma^2-\omega_0^2},
$
which allows for selectively removing it through mode-bypassing protocol, as  introduced in Eq.~(\ref{mode_bypassing}). In Ref.~\cite{bagui_2025}, it was shown that the operator Mpemba effect is absent for unital dynamics when the acceleration protocol is implemented through a unitary transformation. In contrast, we show here that a genuine operator Mpemba effect can emerge under the same unital dynamics through the mode-bypassing procedure. 

To illustrate this point, we consider a generic single-qubit operator
$
\mathcal{O}=b_0\mathbb{I}+\vec{b}\cdot\vec{\sigma},
$
and construct the transformed operator $\widetilde{\mathcal{O}}$ by eliminating its overlap with the SDM. As a consequence, the transformed operator relaxes on a faster timescale governed by the next slowest eigenmode. 
This behavior is shown in Fig.~(\ref{Mpemba_unital}). Fig~\ref{Mpemba_unital}(a) shows the evolution of the dressed distance, where the transformed operator $\widetilde{\mathcal{O}}$ initially lies farther from the steady state but converges more rapidly than the original operator $\mathcal{O}$. Fig.~\ref{Mpemba_unital}(b) shows the corresponding expectation values of the operator, before and after transformation, and clearly highlighting the accelerated relaxation induced by the mode-bypassing protocol.

Interestingly, in contrast to the unital map with $\sigma_x$ dissipator, if we consider the single qubit setup evolving under a Davies map \cite{PhysRevLett.133.140404,davies1979generators} where the $\sigma_{+}$ and $\sigma_{-}$ are the Lindblad dissipators with rates satisfying the detailed balance condition, the operator Mpemba effect is completely absent (for details see Ref.~\cite{supp}). A similar absence of the Mpemba effect was recently reported in Ref.~\cite{bagui_2025}, where accelerated relaxation was induced through a unitary transformation.

\vspace{0.3em}

\textit{Example 2: Non-equilibrium XX-spin chain.--}
\begin{figure}[h!]
    \centering
\includegraphics[width=\columnwidth]{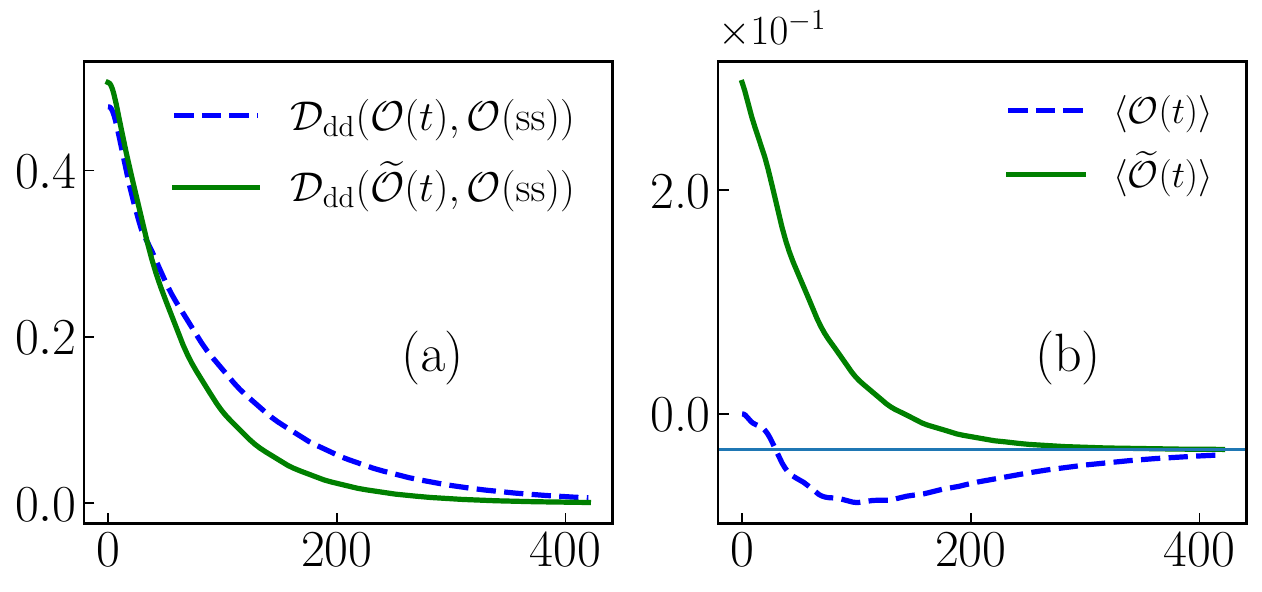}
\includegraphics[width=\columnwidth]{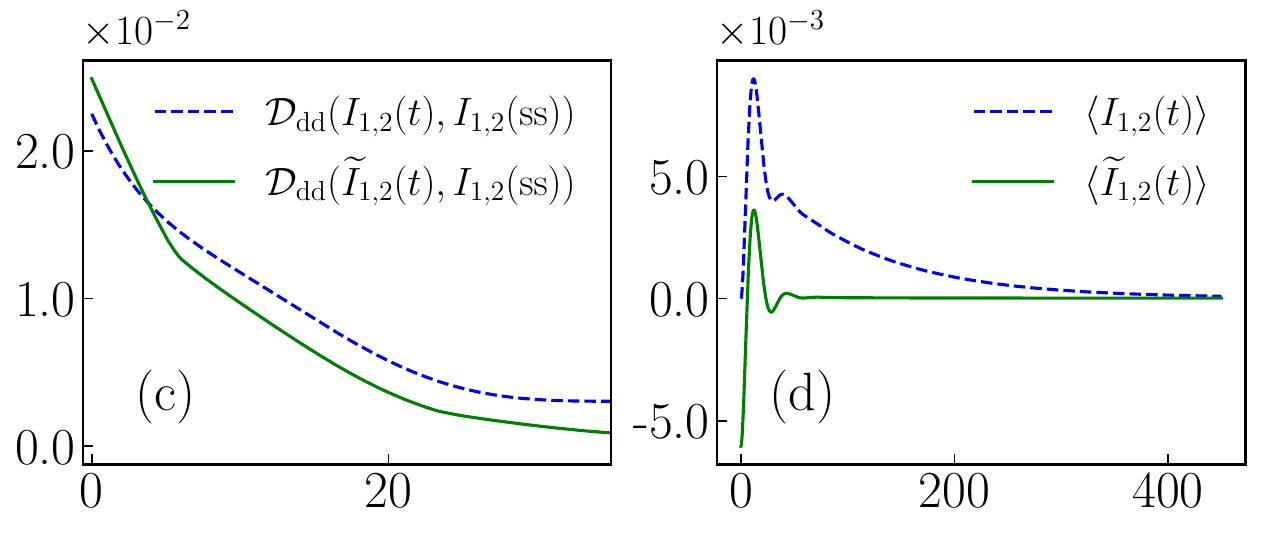} 
     
\includegraphics[width=\columnwidth]{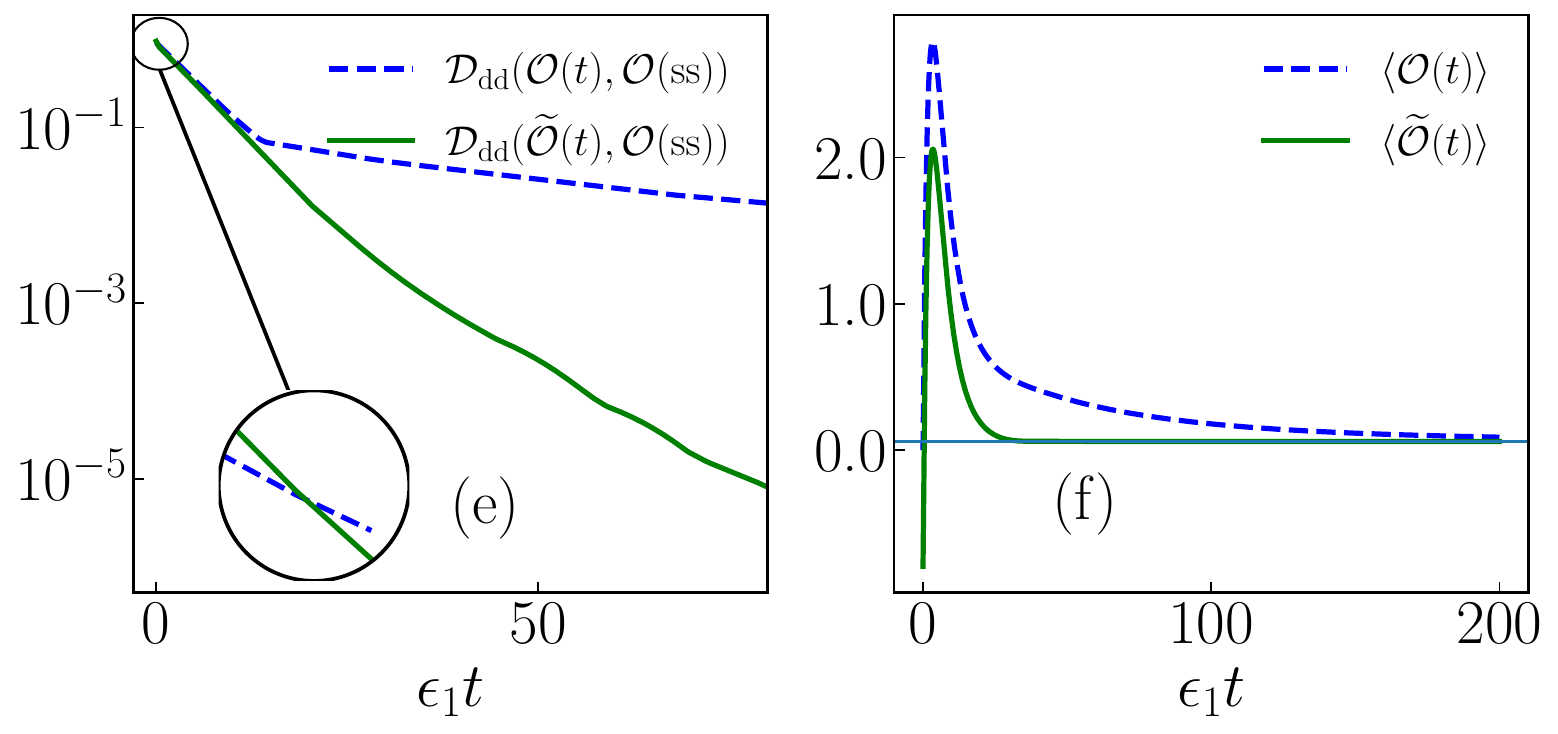}
    \vspace{-0.2cm}
\caption{Results for the Non-equilibrium $XX$-spin chain: Demonstration of the operator Mpemba effect for operators with different degree of locality for the boundary-driven spin chain with $N=4$. (a) Local operator: population imbalance $\mathcal{O}=n_2-n_3$. The transformed operator $\widetilde{\mathcal{O}}$ exhibits faster convergence to the steady state despite  initially larger dressed distance.
(b) Dynamics for the corresponding expectation values starting with the initial state $\ket{\psi_0}=\ket{1}^{\otimes 4}$, ($\ket{1}$ is the spin down state) showing accelerated relaxation and convergence to the same steady state.
(c) Two-site operator: current $I_{1,2}$. The transformed operator $\widetilde{I}_{1,2}$ relaxes exponentially faster than $I_{1,2}$, despite a larger initial dressed distance.
(d) Dynamics for the corresponding expectation values starting with the initial state $\ket{\psi_0}=\ket{0}^{\otimes 4}$, ($\ket{0}$ is the spin up state), confirming faster relaxation to the same steady state.
(e) Non-local operator: $\mathcal{O} = \mathbb{I}_1 \otimes \sigma_z^{(2)} \otimes \sigma_z^{(3)} \otimes \sigma_z^{(4)} - \sigma_z^{(1)} \otimes \sigma_z^{(2)} \otimes \sigma_z^{(3)} \otimes \mathbb{I}_4.$ Although $\widetilde{\mathcal{O}}$ is far from the steady state, it relaxes faster due to vanishing overlap with the five dominant SDMs. The inset shows the zoomed version of the initial distance.
(f) Corresponding expectation values for the initial state $\ket{\psi_0}=\ket{1}^{\otimes4}$, showing accelerated convergence.
For simulation, we use the parameters $g=0.05$, $\gamma_1=0.2$, $\gamma_N=0.5$, 
$\epsilon_{1,2,3,4}=(1,0.8,0.6,0.5)$, $T_1=1.5$, $T_N=1$, and $\mu_1=\mu_N=0$. }
    \label{XX_chain}
\end{figure}
As a second example, we consider a boundary-driven $XX$ spin chain described by the Hamiltonian $H=\sum_{i=1}^{N}\epsilon_i \, \sigma_{+}^{(i)}\sigma_{-}^{(i)}
+ g\sum_{i=1}^{N-1} 
\big( \sigma_{+}^{(i)}\sigma_{-}^{(i+1)}
+ \sigma_{-}^{(i)}\sigma_{+}^{(i+1)} \big)$, where $\epsilon_i$ is the onsite energy, $g$ is the nearest-neighbor hopping strength and $N$ is the number of qubits. 
The dynamics of the setup is governed by Eq.~(\ref{lindblad_state}), with jump operators corresponding to two boundary sites are given by $L_i^{\pm}=\sigma_{\pm}^{(i)}$, $i\in\{1,N\}$, and associated jump rates are $\gamma_i f_i(\epsilon_i)$ and $\gamma_i[1-f_i(\epsilon_i)]$, respectively. Here 
$f_i(\epsilon_i)$ is the Fermi distribution function evaluated at the onsite energy $\epsilon_i$ with temperature $T_i$ and chemical potential $\mu_i$, which are fixed at different values at the two edges. As a result, the system is driven out-of-equilibrium and the setup possesses a unique current carrying steady-state. Within this setup, we investigate the emergence of the genuine Mpemba effect for the operators using the mode-bypassing procedure. 

We consider three representative operators with an increasing degree of non-locality. Here we present results for $N=4$. First, as a local (single-site) observable, we study the population imbalance between the consecutive sites, defined as $n_i - n_{(i+1)}$ with $n_i=(\mathbb{I}+\sigma_z^{(i)})/2$.
Next, as a two-site observable, we consider bond current which is defined as $I_{i,i+1}=-ig\big[\mathbb{I}_1\otimes..\otimes \big(\sigma_{-}^{(i)}\otimes \sigma_{+}^{(i+1)}-\sigma_{+}^{(i)}\otimes \sigma_{-}^{(i+1)}\big)\otimes \mathbb{I}_{i+2}...\otimes \mathbb{I}_{N}\big].$
Finally, as a more non-local (three-site) observable, we analyze a string operator of the form
$
\mathcal{O} = \mathbb{I}_1 \otimes \sigma_z^{(2)} \otimes \sigma_z^{(3)} \otimes \sigma_z^{(4)} - \sigma_z^{(1)} \otimes \sigma_z^{(2)} \otimes \sigma_z^{(3)} \otimes \mathbb{I}_4.
$

In Fig.~(\ref{XX_chain}), we demonstrate the emergence of the operator Mpemba effect for these different classes of observables. For the local operator corresponding to imbalance, the transformed operator exhibits faster relaxation compared to the original one, despite being initially farther from the steady state in the dressed distance measure, as shown in Fig~(\ref{XX_chain}) (a)-(b). A similar behavior is observed for the two-body current operator $I_{1,2}$, where bypassing the SDM leads to a clear exponential speedup in the relaxation dynamics, as clearly demonstrated in Fig~(\ref{XX_chain}) (c)-(d). The effect persists even for the non-local three-site string operator. Here, we bypass the five dominant slow decay modes of the initial observable using the multimode bypassing procedure. Although the transformed observable is initially farther from the steady state, it nevertheless reaches the steady state more rapidly than the original observable, thereby exhibiting the operator Mpemba effect, as clearly shown in Fig~(\ref{XX_chain}) (e)-(f).
In all these cases, we verify the validity of the bound in Eq.~\eqref{inequality} and present results in \cite{supp}.

\vspace{0.3em}
\textit{Example 3: One-dimensional bosonic lattice with bulk dephasing.-- }
\begin{figure}
    \centering
    \includegraphics[width=\columnwidth]{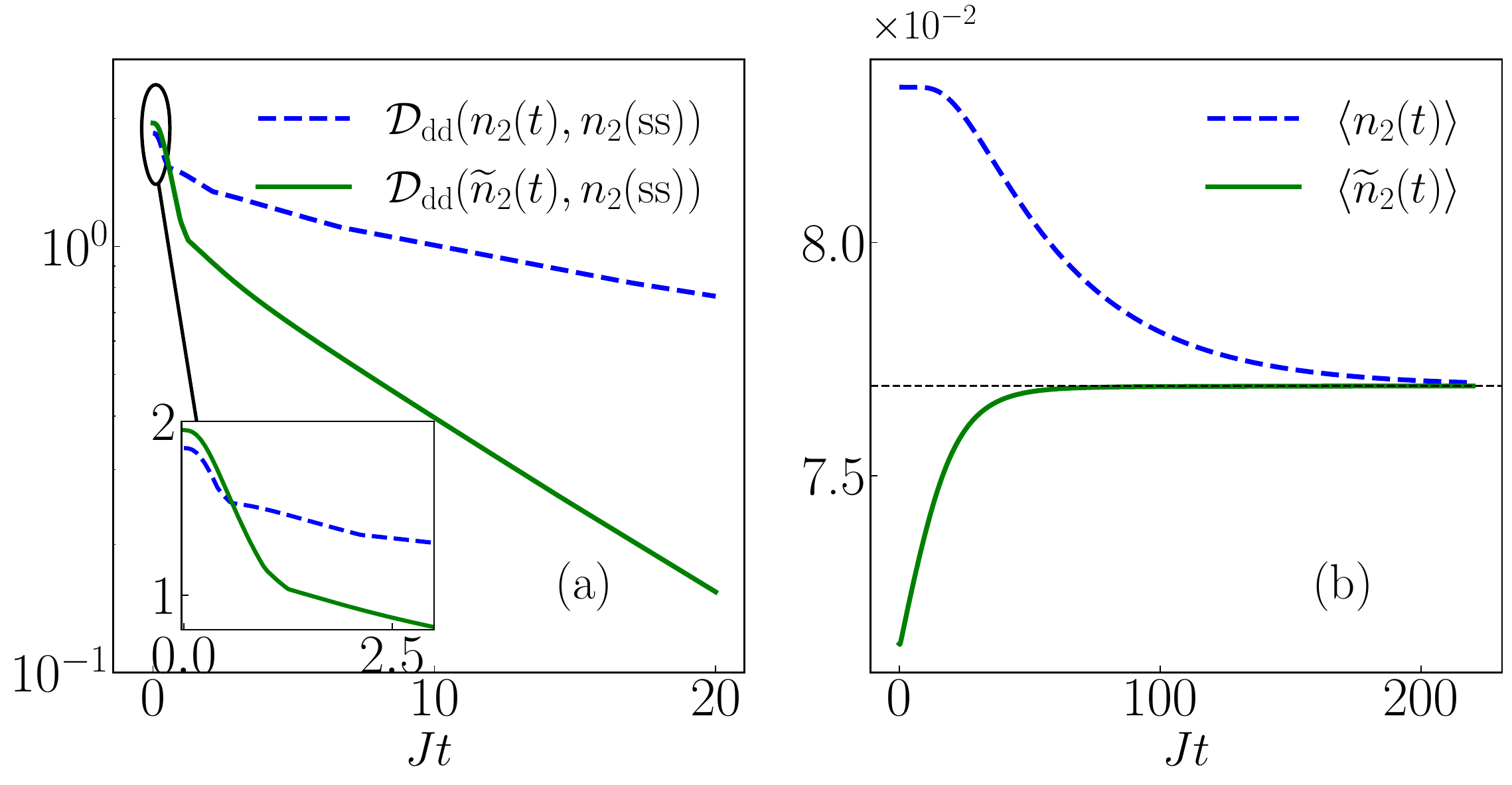}
    \caption{Results for one-dimensional bosonic lattice subjected to dephasing: (a) The transformed operator $\widetilde{n}_2(t)$ initially lies farther from the steady state value $n_2^{(\mathrm{ss})}$ in terms of dressed distance (for this case it reduces to trace distance, as the map is unital), yet it relaxes more rapidly than the local number density $n_2$ at the second site. In contrast, $n_2$, although closer to equilibrium at initial times, exhibits slower relaxation. This counterintuitive behavior highlights the presence of the genuine Mpemba effect for operators.
(b) The expectation value evaluated with respect to the initial state $\rho_0 = \frac{1}{L-1} \sum_{i=1}^{L-1} \ket{i}\bra{i}$ further confirms the accelerated relaxation of the transformed operator. The simulations are performed for parameters $L = 11$, $J = 1$, and $\gamma = 5$.
}
\label{dephasing_local}
\end{figure}
As a final example, we consider a one-dimensional (1D) bosonic lattice of $L$ coupled sites with nearest-neighbor tunneling. The Hamiltonian for the lattice is given as $
H = \sum_{n=1}^{L-1} J \left( a_{n}^{\dagger} a_{n+1} + \mathrm{h.c.} \right)
$. Here, $J$ denotes the tunneling amplitude between adjacent sites, and $a_n^{\dagger}$ ($a_n$) are the bosonic creation (annihilation) operators at site $n$. The lattice is further subjected to uniform onsite dephasing \cite{Longhi:24,Liang_2025} with dephasing strength $\gamma_i = \gamma$, and the jump operator is given as $L_n = a_n^{\dagger} a_n$, $n=1, 2, \cdots L$.
For this setup, we investigate the presence of genuine operator Mpemba effect by focusing on the single particle sector under strong dephasing limit $\gamma \gg J$ \cite{Longhi:24} and intermediate dephasing limit $\gamma \sim J$. Here, we discuss the results in the strong-dephasing limit, and the results for moderate dephasing, $\gamma \sim J$, are presented in Ref.~\cite{supp}.

In the strong dephasing limit, the coherences in the site basis are strongly suppressed, leading to the evolution confined entirely in the population sector $p_n=\rho_{n,n}(t)$. The population dynamics can be described by a Pauli master equation \cite{breuer2002theory,Longhi:24,Longhi_prl_2024} $\frac{d\mathbb{P}}{dt}=\mathcal{W}\,\mathbb{P}$ where $\mathbb{P}=(p_1, p_2, \cdots, p_L)^T$.
 Since the rate matrix $\mathcal{W}$ is Hermitian, its left and right eigenmatrices are identical. These eigenmodes govern the relaxation dynamics both for the operators and states and play the central role in the Mpemba effect. For a lattice of size $L$, the  SDM of $\mathcal{W}$ takes the form
$
r_1=\sum_{n=1}^{L}\Phi_1(n)\ket{n}\bra{n},
$
\cite{Longhi:24} where $\Phi_1(n)=\sqrt{\frac{2}{L}}\cos\!\left[\frac{\pi(2n-1)}{2L}\right]$, and $n$ labels the lattice sites. To reveal the operator Mpemba effect, we analyze the dynamics of the local observable $n_i = a_i^{\dagger} a_i$ and its transformed counterpart $\widetilde{n}_i$, constructed via the mode-bypassing protocol, as given in Eq.~(\ref{mode_bypassing}).

As shown in Fig.~(\ref{dephasing_local})(a), the bare observable $n_2$ exhibits slow relaxation towards its steady-state value $n_2^{(\mathrm{ss})}$, governed by its finite overlap with the SDM $r_1$. In contrast, the transformed operator $\widetilde{n}_2$ is explicitly engineered to eliminate this overlap. As a result, it bypasses the SDM and relaxes significantly faster, despite being initially farther from the steady state. This effect is demonstrated in Fig.~(\ref{dephasing_local})(b), where $\langle \widetilde{n}_2(t) \rangle$ displays a clear exponential speed-up compared to $\langle n_2(t) \rangle$. This provides direct evidence of a genuine operator Mpemba effect driven by mode bypassing. Crucially, the transformation preserves the steady state, ensuring that both observables relax to the same asymptotic value while following markedly different dynamics. We further show in \cite{supp} that bypassing multiple slow decay modes leads to an even stronger enhancement of the relaxation dynamics, resulting in a further reduction of the relaxation timescale.

\textit{Summary.--}
We investigate the emergence of a genuine Mpemba effect for operators in Markovian open quantum systems. Unlike quantum states, operators evolve under the adjoint Liouvillian and therefore do not follow a CPTP map. Exploiting this distinction, we introduce a systematic procedure to accelerate operator relaxation by selectively removing the slowest decaying mode, while preserving convergence to the same steady-state value. Furthermore, we derive a quantitative bound on the change in the dressed distance induced by this mode-bypassing transformation. Our results are particularly relevant in scenarios where the rapid equilibration of specific observables is more important than the relaxation of the full quantum state.
Our work thus establishes that the quantum Mpemba effect is not restricted to the relaxation of quantum states, but can also occur directly at the level of physical observables.  This broadens the conceptual scope of anomalous relaxation phenomena in quantum systems and opens promising future directions that includes extending the present analysis for time-dependent systems and for systems evolving under non-Markovian dynamics. 

\vspace{0.1em}
\textit{Acknowledgements.--}
PB acknowledges the University  Grants Commission (UGC), Government of India, for the research fellowship (Ref No.- 231620073714). BKA acknowledges CRG Grant No. CRG/2023/003377 from Science and Engineering Research Board (SERB), Government of India. 

\bibliography{bibliography}

\setcounter{equation}{0}
\setcounter{figure}{0}
\renewcommand{\theequation}{S\arabic{equation}}
\renewcommand{\thefigure}{S\arabic{figure}}

\onecolumngrid
\newpage

\begin{center}
{\textbf{\large{Supplemental Material: Quantum Mpemba effect for operators in open systems}}}
\end{center}

\section{Nature of transformed operators under Mode Bypassing}
\label{AppendixA}
In this section, we discuss how the mode-bypassing protocol can transform an initially local operator into a nonlocal one. The relaxation dynamics of an operator is governed by its decomposition in the eigenmodes of the underlying dynamical generator, which in this case is the adjoint Liouvillian $\mathcal{L}^{\dagger}$. Any operator $\mathcal{O}$ with a finite overlap with its slowest decay mode (SDM) decides its relaxation timescale in the long-time limit. Therefore, controlling this overlap provides a direct route to modify the relaxation behavior. In particular, the mode-bypassing procedure introduced in Eq.~(\ref{mode_bypassing}) suppresses the contribution of the SDM, thereby accelerating the relaxation of the transformed operator. However, this acceleration comes at a cost. If one starts from a local operator, the application of the mode-bypassing transformation generally produces a nonlocal operator. This originates from the fact that the right and left eigenvectors of the Liouvillian $\mathcal{L}$ are, in general, nonlocal when expressed in the Pauli operator basis. Consequently, eliminating the overlap with the SDM inevitably introduces nonlocal components into the transformed operator. 

To illustrate this explicitly, we consider Example~2 from the main text, namely the boundary-driven XX spin chain. Considering the initial operator to be the local number operator at the second site, $\mathcal{O}=(\mathbb{I}+\sigma_z^{(2)})/2$, we find that applying the mode-bypassing procedure to eliminate its overlap with the SDM and produces a transformed operator that is manifestly nonlocal in Pauli basis. As shown in Fig.~\ref{nonlocality}, although the original operator has support only on a single site, the transformed operator develops finite contributions from both local and increasingly nonlocal operator sectors. In particular, the weight of multi-site operator components, which we define as 
\begin{equation}
\label{coefficient}
|c|=\mathrm{Tr}(\widetilde{\mathcal{O}}\,\sigma_i^{\otimes4})/\mathrm{Tr}[r_1\mathcal{O}]
\end{equation}
clearly demonstrates that the elimination of the SDM generates extended correlations across the system. Here $\sigma_i\in[\mathbb{I}_2,\sigma_x\sigma_y,\sigma_z]$.

It is important to emphasize that the behavior of the transformed operator depends on the structure of the SDM itself. For example, as shown in Example~3 in the main text, under the strong dephasing limit ($\gamma \gg J)$ the SDM is local in the number basis. Consequently, bypassing the SDM in that case does not introduce non-locality into the transformed operator, as a result accelerated or transformed operator remains local.

\begin{figure}[h!]
    \centering
\includegraphics[width=0.75\columnwidth]{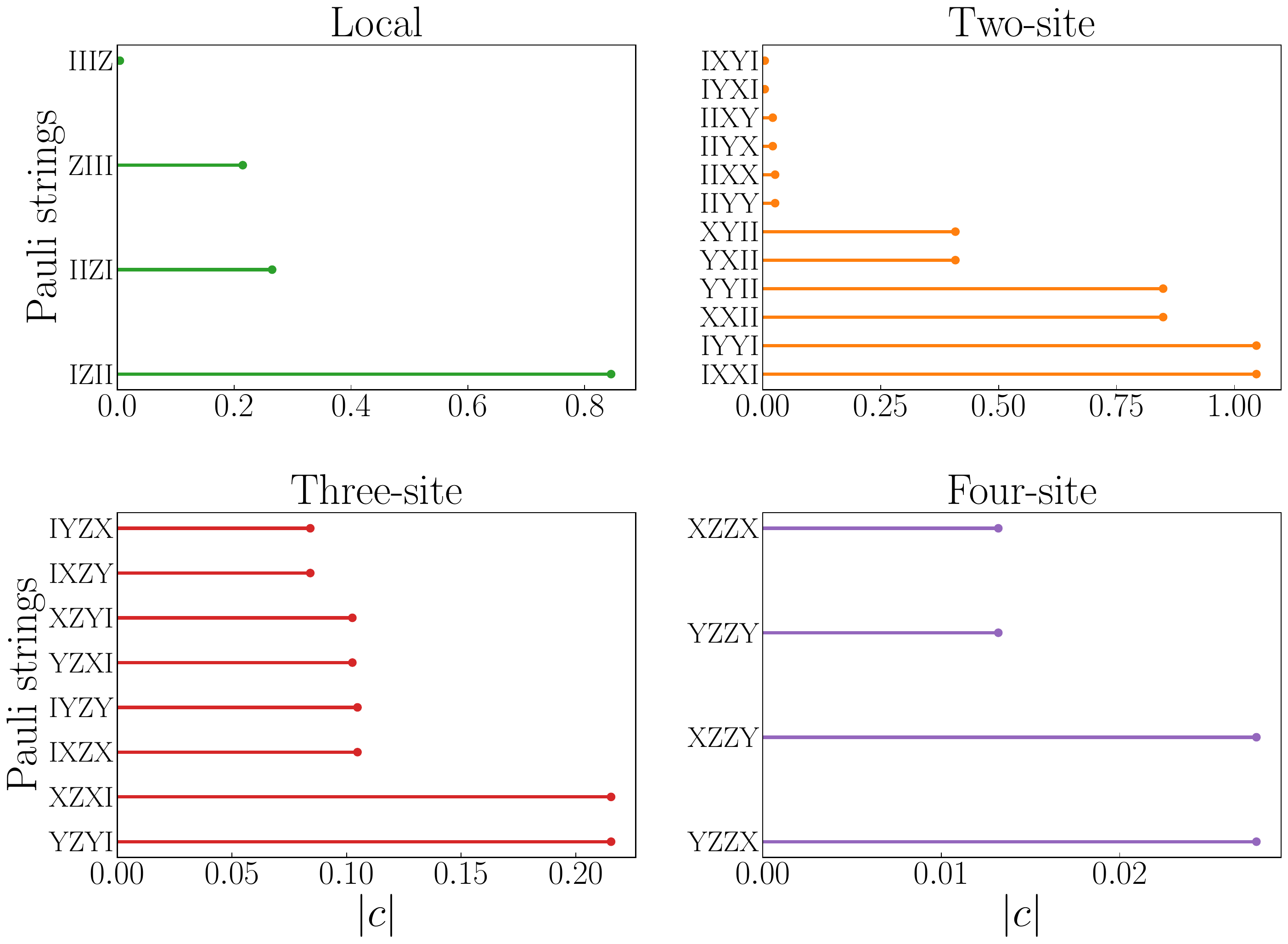}
    \caption{ Plot to show the nonlocal behaviour of the transformed operator generated by the mode-bypassing procedure in the non-equilibrium $XX$-spin chain model [Example 2 of the main text], starting from a local operator $\mathcal{O}=(\mathbb{I}+\sigma_z^{(2)})/2$. The figure shows the distribution of the magnitudes of the Pauli decomposition coefficients, $|c|$, as defined in Eq.~\eqref{coefficient} of the transformed operator, grouped according to their Pauli weight (from 1-site to 4-site sectors). Each panel corresponds to a fixed locality sector and displays the coefficients associated with Pauli strings belonging to that sector. Although the initial operator is strictly local, the transformed operator develops finite contributions from increasingly nonlocal sectors after removing the slowest decay mode. The coefficients are normalized by the overlap of the initial operator with the right eigenvector $r_1$ of $\mathcal{L}$, corresponding to $\lambda_1$ and only terms satisfying $|c|>10^{-4}$ are retained. The parameters used are $g=0.05$, $\gamma_1=0.2$, $\gamma_N=0.5$, $\epsilon_{1,2,3,4}=(1,0.8,0.6,0.5)$, $T_1=1.5$, $T_N=1$, and $\mu_1=\mu_N=0$.
    }
\label{nonlocality}
\end{figure}

\subsection{Experimental strategies to measure the transformed operator}
In this subsection,  we discuss possible experimental strategies to measure the expectation value of the transformed operators which are generally non-local. In particular, we introduce two different approaches that allow one to access this quantity in practice.

\vspace{0.4em}
\textit{\textbf{Method 1.--}}
Recall that, the transformed operator is defined as
$
    \widetilde{\mathcal{O}} = \mathcal{O} - \mathrm{Tr}(r_1 \mathcal{O})\, l_1,
$
where $r_1$ and $l_1$ denote the right and left eigenoperators corresponding to the SDM of the operator $\mathcal{O}$. This transformation can be interpreted as projecting $\mathcal{O}$ out of the slowest decay mode (SDM) using the projector
$\Pi = \mathbb{I} - \ket{l_1}\bra{r_1}.
$, where $\mathbb{I}$ is the Identity operator.  As a result, the transformed operator $\widetilde{\mathcal{O}}$ relaxes with a faster timescale $\tau_2 = 1/|\mathrm{Re}(\lambda_2)|$, where $\lambda_2$ is the next slowest decay mode of the operator $\mathcal{O}$. The expectation value of the transformed operator evolves as
\begin{equation}
\mathrm{Tr}\big(\widetilde{\mathcal{O}}(t)\rho_0\big)
= \mathrm{Tr}(\rho_t \mathcal{O})
- \mathrm{Tr}(r_1 \mathcal{O})\, \mathrm{Tr}(l_1 \rho_0)\, e^{-\lambda_1 t}.
\end{equation}
This expression provides a practical route for experimental implementation. Instead of directly measuring the transformed operator $\widetilde{\mathcal{O}}$, one can measure the expectation value of the original operator $\mathcal{O}$ and subtract the known contribution from the SDM \cite{chatterjee_2025_direct}.

\vspace{0.4em}
\textit{\textbf{Method 2.--}}
We next discuss an ancilla-based technique to measure the expectation value of the transformed operator $\widetilde{\mathcal{O}}(t)$.
We introduce an ancilla qubit prepared in the superposition state $ \ket{\psi^{\mathrm{anc}}} =\frac{\ket{0} + \ket{1}}{\sqrt{2}},$
with corresponding density matrix, given as $\rho_A = \ket{\psi^{\mathrm{anc}}}\bra{\psi^{\mathrm{anc}}}$. Just before the measurement $(t^{-})$, the total system-ancilla state is given as a product state i.e., 
$\rho_{\mathrm{tot}}(t) = \rho_A \otimes\rho_s(t).
$
One can then introduce an interaction between the system and the ancilla governed by the Hamiltonian
$H_{\mathrm{int}} = g\, \sigma_z^{\mathrm{anc}} \otimes \widetilde{\mathcal{O}},$
applied for a duration $\tau$. Here $g$ is the interaction strength. The corresponding unitary evolution is given by
\begin{equation}
    U = e^{-i g \tau \sigma_z \otimes \widetilde{\mathcal{O}}}
    = \ket{0}\bra{0} \otimes e^{-i g \tau \widetilde{\mathcal{O}}}
    + \ket{1}\bra{1} \otimes e^{i g \tau \widetilde{\mathcal{O}}}.
\end{equation}
After the interaction, the total state becomes
$
    \rho_{\mathrm{tot}}(t^{+}) = U \big( \rho_A \otimes \rho_s(t) \big) U^\dagger.
$
The reduced state of the ancilla is obtained by tracing out the system i.e.,
$\rho_A(t^{+}) = \mathrm{Tr}_s \big[ \rho_{\mathrm{tot}}(t^{+}) \big].$
The off-diagonal element of the ancilla is given by
\begin{equation}
    \bra{0}\rho_A(t^{+})\ket{1}
    = \frac{1}{2} \mathrm{Tr}_s \big[ \rho_s(t)\, e^{-2 i g \tau \widetilde{\mathcal{O}}} \big].
\end{equation}
In the weak-coupling and short interaction time limit, i.e., when $g\tau \ll 1$, one can write 
$
    e^{-2 i g \tau \widetilde{\mathcal{O}}}
    = \mathbb{I} - 2 i g \tau \widetilde{\mathcal{O}} + \mathcal{O}((g\tau)^2).
$
Using this expansion, one finds that the expectation value of the ancilla observable $\sigma_y$ is directly proportional to the expectation value of the transformed operator
\begin{equation}
\langle \sigma_y(t+\tau) \rangle = 2 g \tau \, \langle \widetilde{\mathcal{O}}(t) \rangle.
\end{equation}
This relation provides an operational protocol to measure the expectation value of the transformed operator $\widetilde{\mathcal{O}}$ indirectly via the ancilla \cite{Bose_kim_Prl_2016,PhysRevLett.112.190402}, avoiding the need for direct measurement of a generic non-local operator.

\section{Proof for the bound in Eq.~(9)}
\label{AppendixB}
In this section, we aim to show that for any arbitrary unitary operator $U$,
\begin{equation}
\|X\| = \mathrm{Tr}(U_X^\dagger X) \ge \mathrm{Re}\,\big[\mathrm{Tr}(U^\dagger X)\big],
\end{equation}
with equality holds if and only if $U = U_X$. This condition was used in the main text to derive the lower and upper bounds for the dressed distance, as given in Eq.~\eqref{inequality}. Note that $X = U_X |X|$ is the polar decomposition for $X$, and $|X| = \sqrt{X^\dagger X} \ge 0$. Following the polar decomposition, we can write
\begin{equation}
\|X\| = \mathrm{Tr}|X| = \mathrm{Tr}(U_X^\dagger X).
\end{equation}
For an arbitrary unitary operator $U$, we define $V = U^\dagger U_X$, which is also a  operator. As a result, we can write 
\begin{equation}
\mathrm{Re}\,\big[\mathrm{Tr}(U^\dagger X)\big]
= \mathrm{Re}\,\big[\mathrm{Tr}(U^\dagger U_X |X|)\big]
= \mathrm{Re}\,\big[\mathrm{Tr}(V |X|)\big].
\end{equation}
Let us now consider the spectral decomposition of $|X|$ and write 
$
|X| = \sum_i s_i \ket{i}\bra{i}$, with $s_i \ge 0$.
Then, we can write 
\begin{equation}
\mathrm{Re}\,\big[\mathrm{Tr}(V|X|)\big]
= \sum_i s_i\, \mathrm{Re}\,\big[\bra{i}V\ket{i}\big].
\end{equation}
Employing the Cauchy--Schwarz inequality, we write
\begin{equation}
|\langle{i}|j\rangle|^2\leq |\langle i|i\rangle|^2 |\langle j|j\rangle|^2=1,
\end{equation}
where $\ket{j}=V\ket{i}$ and as $V$ is a unitary operator, it does not change the norm of the vector $\ket{i}$.
Now, for any complex number $z$, $\mathrm{Re} \big[z \big] \le |z|$, we can write 
\begin{equation}
\mathrm{Re}\,\big[\bra{i}V\ket{i}\big] \le 1.
\end{equation}
Therefore,
\begin{equation}
\|X\| =\mathrm{Tr}|X|= \sum_i s_i \ge
\mathrm{Re}\,\big[\mathrm{Tr}(U^\dagger X)\big],
\end{equation}
with equality holds if and only if $\mathrm{Re}\,\big[\bra{i}V\ket{i}\big] = 1$ for all $i$, which implies $\bra{i}V\ket{i} = 1$ for all $i$. This is possible only when $V$ is an identity operator, and hence $U = U_X$.

\section{Enhanced Relaxation for Operators via Multiple Mode-Bypassing}
\label{AppendixC}
\begin{figure}
    \centering
    \includegraphics[width=0.75\columnwidth]{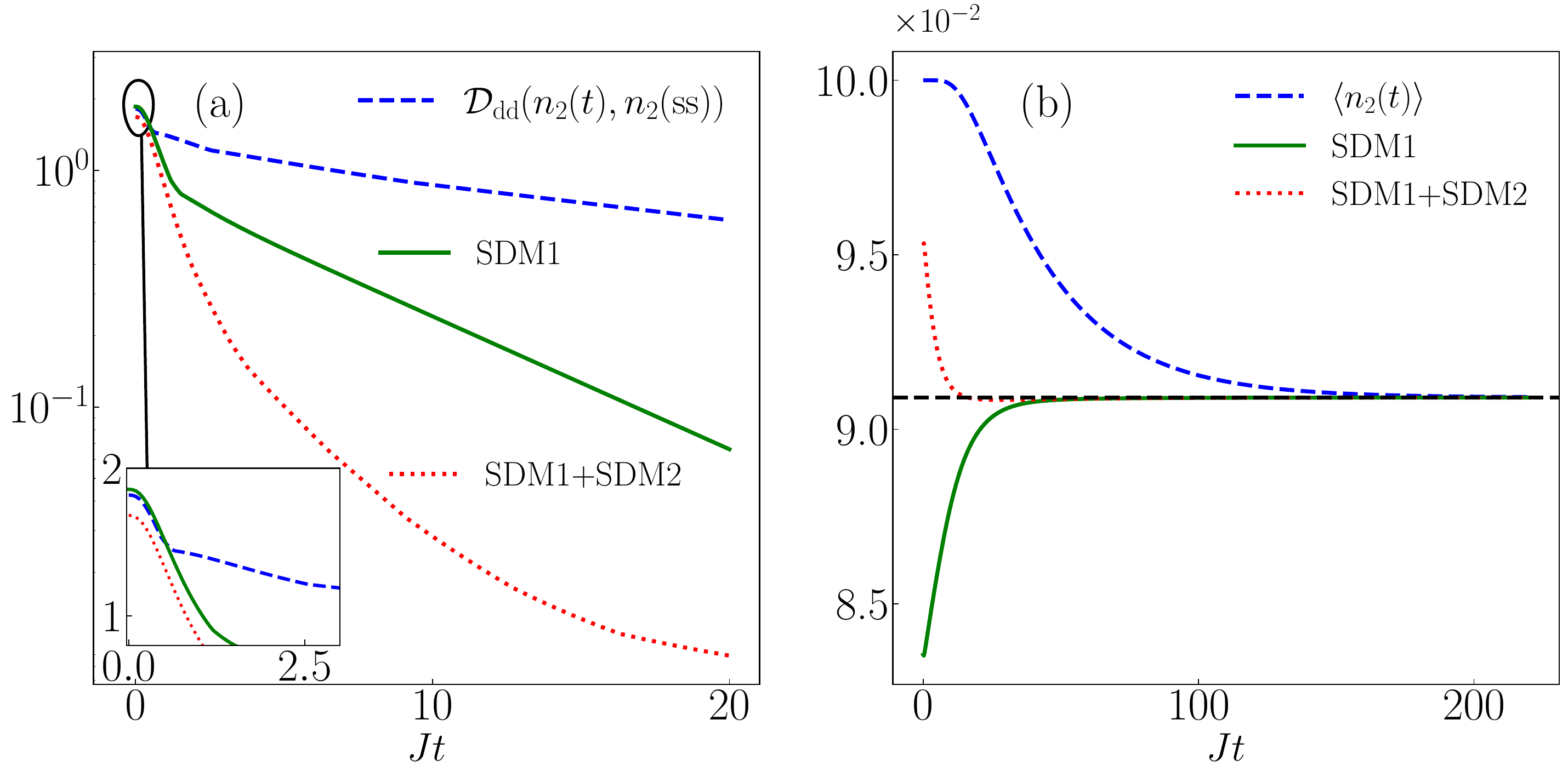}
    \caption{
Enhanced relaxation via successive mode-bypassing. 
(a) Dressed distance dynamics displaying the relaxation dynamics for three cases: the original operator, taken as the local site density $(n_2)$ (blue dashed), the transformed operator obtained by bypassing the slowest decay mode (SDM1) (green solid), exhibiting the Mpemba effect (see inset), and the further transformed operator obtained by bypassing the two slowest modes (SDM1 and SDM2) (red dotted), demonstrating enhanced relaxation. 
(b) Corresponding dynamics for operator expectation values evaluated with respect to the initial state 
$\rho_0 = \frac{1}{L-1}\sum_{i=1}^{L-1} |i\rangle \langle i|$, 
showing progressively faster convergence of the local density $n_2$ to the same steady state. For the numerics, we consider the parameters $L=11$, $J=1$, and the dephasing strength $\gamma=5$.
}
\label{enhanced_relaxation}
\end{figure}
In the main text, we established that the accelerated convergence of the transformed operator $\widetilde{\mathcal{O}}$ to the steady state arises by eliminating the contribution of the SDM of the original operator $\mathcal{O}$. This procedure can be systematically extended to further accelerate the dynamics by successively bypassing multiple slow decay modes. To illustrate this explicitly, we revisit Example 3 of the main text, which is a one-dimensional bosonic setup subjected to dephasing. We showed in the main text that, starting from the population density of a local site $(n_i)$ as the initial operator, we bypass the contribution of its SDM and observe the emergence of the Mpemba effect for operators. One can however proceed further by bypassing other slow decaying modes, which can lead to a further acceleration of the relaxation towards the asymptotic value.

In Fig.~\ref{enhanced_relaxation}(a) and (b), we demonstrate the enhanced accelerated relaxation where we bypass not only the slowest decay mode, which we call SDM1, but also the next slow decay mode, SDM2. In Fig.~~\ref{enhanced_relaxation}(a), we plot the dynamics of the dressed distance for the local density operator. We observe a significant enhancement in relaxation upon bypassing the two slowest decay modes (SDM1 and SDM2), as shown by the red dotted curve, compared to the relaxation of the original operator (blue dashed line), and the transformed operator with only the slowest decay mode (SDM1) bypassed (green solid line). The associated faster convergence in the expectation value of the operator to the same steady state is observed in Fig.~\ref{enhanced_relaxation}(b).

\section{Absence of operator Mpemba effect in single qubit evolving under Davies map}
\label{AppendixD}
In this section, we show that for an arbitrary operator in the single-qubit Hilbert space, evolving under the Davies map, there can not be Mpemba effect for operators. Any general operator in the single-qubit Hilbert space can be written as 
$\mathcal{O} = b_0 \, \mathbb{I} + \vec{b} \cdot \vec{\sigma}$, 
where $\vec{b} \in \mathbb{R}^3$ is a real three-dimensional vector and 
$\vec{\sigma} = (\sigma_x, \sigma_y, \sigma_z)$ is the vector of Pauli matrices. 
The single-qubit system is described by the Hamiltonian: $H = \omega_0 \, \sigma_z/2$,
where $\sigma_z$ is the Pauli-Z operator and $\omega_0$ is the energy gap between the two levels. We choose the jump operators as $\sigma_+=\Ket{1}\Bra{0}$ and $\sigma_-=\sigma_{+}^{\dagger}$ with corresponding rates $\gamma n_b$ and $\gamma(1 + n_b)$, respectively. Here $n_b=1/(\exp(\hbar\omega_0/k_BT)-1)$ is the Bose-Einstein occupancy factor at bath temperature $T$, and $k_B$ is the Boltzmann constant.  In this case,  the eigenvalues $\lambda_1$ and $\lambda_2$ of the adjoint Liouvillian form a complex-conjugate pair ($\lambda_2=\lambda_1^{\star}$) corresponding to the slowest decay mode (with equal real parts). The associated right eigenmatrices $r_1$ and $r_2$ are complex conjugates of each other and are purely off-diagonal, a characteristic feature of the Davies map structure. Therefore, to 
suppress the contribution of these slowest decaying modes and thereby achieving accelerated relaxation in the operator dynamics, 
one can perform a mode-bypassing procedure on $\mathcal{O}$ such that it becomes diagonal. 
The transformed operator then takes the form 
$\widetilde{\mathcal{O}} = \begin{pmatrix} b_0 + b_z & 0 \\ 0 & b_0 - b_z \end{pmatrix}$
where $b = \sqrt{b_x^2 + b_y^2 + b_z^2}$. As a result, the operator will quickly relax to the steady state. To investigate how the initial dressed distance relative to the steady state changes after the mode bypassing operation, we compute the dressed distance for the two cases. In our setup, the steady state is a detailed-balance thermal state, given by $\rho_{\mathrm{ss}} = e^{-\beta H}/\mathcal{Z}$, where $H=\frac{\omega_0}{2} \sigma_z$, $\beta = 1/(k_B T)$ and the partition function $\mathcal{Z} = 2\cosh{(\beta \omega_0 / 2)}$.  The initial dressed distance can be written as  
$
\mathcal{D}_\mathrm{dd}(\mathcal{O}, \mathcal{O}_{\mathrm{ss}}) = \sum_{i=1}^{2} |\lambda_i'|
= 2\sqrt{b_x^2 + b_y^2 + (2b_z/\mathcal{Z})^2}/\mathcal{Z},
$
where $\lambda_i'$ are the eigenvalues of the operator 
$\rho_{\mathrm{ss}}^{1/2} (\mathcal{O} - \mathcal{O}_{\mathrm{ss}}) \rho_{\mathrm{ss}}^{1/2}$.  
After the transformation, the distance between the transformed operator and the steady state becomes  
$
\mathcal{D}_\mathrm{dd}(\widetilde{\mathcal{O}}, \mathcal{O}_{\mathrm{ss}}) 
= \sum_{i=1}^{2} |\widetilde{\lambda}_i'|
= 4b_z/\mathcal{Z}^2,
$
which is always smaller than the dressed distance before the mode bypassing operation.
Here, $\widetilde{\lambda}_i'$ denote the eigenvalues of 
$\rho_{\mathrm{ss}}^{1/2} (\widetilde{\mathcal{O}} - \mathcal{O}_{\mathrm{ss}}) \rho_{\mathrm{ss}}^{1/2}$.  
As a result, no Mpemba-like effect for operators can be observed in the single-qubit case with the chosen dissipators.
It is interesting to note that, a similar absence of the Mpemba effect was observed in Ref.~\cite{bagui_2025} where the accelerated relaxation for the operator was obtained via unitary transformation.

\section{Operator Mpemba effect in the intermediate dephasing regime for bosonic lattice}
\label{AppendixE}
In this section, we analyze the dynamics of the one-dimensional bosonic lattice with bulk dephasing, introduced in Example 3 of the main text. We will discuss the regime in which the dephasing strength is comparable to the tunneling amplitude, $\gamma \sim J$ which implies that both coherent tunneling and dissipative processes compete with each other.

Unlike the strong dephasing limit, $\gamma \gg J$ discussed in the main text, the evolution cannot be restricted to the population sector alone in the regime $\gamma \sim J$. Instead, coherences in the site basis remain relevant and are coupled to the populations throughout the dynamics. As a result, the effective classical description in terms of the Pauli master equation is no longer valid, and the full Liouvillian dynamics must be considered. 

To investigate the possibility of operator Mpemba effect in $\gamma \sim J$ regime, we consider the dynamics of the local observable $n_i = a_i^\dagger a_i$ and its mode-bypassed counterpart $\widetilde{n}_i$, constructed using Eq.~(\ref{mode_bypassing}). In Fig~(\ref{dephasing_weak})(a), we demonstrate the occurrence of the Mpemba effect for the operator where crossing is observed in the dynamics of dressed distance (it is proportional to trace distance, as the map here is unital). The corresponding expectation value of the transformed operator captures the faster relaxation (see inset). In Fig.~(\ref{dephasing_weak}) (b), we plot the quantity $\Delta = 
\mathcal{D}_{\mathrm{dd}}(\widetilde{\mathcal{O}}, \mathcal{O}_{\mathrm{ss}})-
\mathcal{D}_{\mathrm{dd}}(\mathcal{O}, \mathcal{O}_{\mathrm{ss}}),
$ which is the difference in the distance value after and before the mode bypassing, as a function of $\gamma$. There is an overall enhancement in the distance measure in $\gamma \sim J$ regimes due to the coherence contribution in the dressed distance compared to the $\gamma\gg J$ region, where coherence is strongly suppressed.

\begin{figure}
    \centering
    \includegraphics[width=0.75\columnwidth]{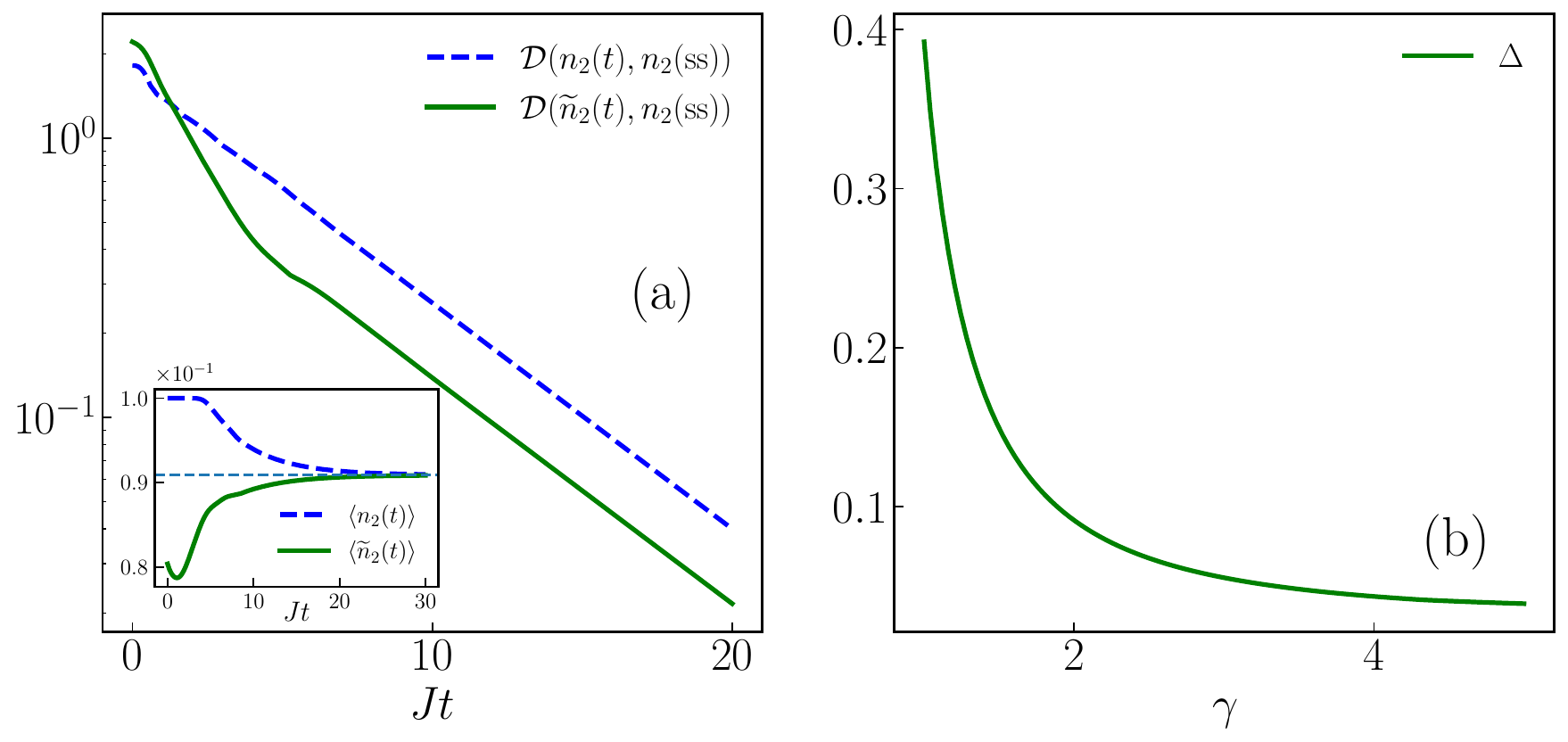}
    \caption{Results for one-dimensional bosonic lattice subjected to dephasing in the $\gamma \sim J$ regime: (a) Plot for the dynamics of dressed distance (in this case it reduces to trace distance, as the map is unital) for the transformed operator $\widetilde{n}_2(t)$ which initially lies farther from the steady-state value $n_2^{(\mathrm{ss})}$, yet it relaxes more rapidly than the local number density of the second site $n_2$. As a result, genuine Mpemba effect for operators can be observed even in the $\gamma\sim J$ regime. The inset shows the accelerated convergence of the expectation value of the transformed operator compared to the original operator.
(b) Plot for the quantity $\Delta = 
\mathcal{D}_{\mathrm{dd}}(\widetilde{\mathcal{O}}, \mathcal{O}_{\mathrm{ss}})-
\mathcal{D}_{\mathrm{dd}}(\mathcal{O}, \mathcal{O}_{\mathrm{ss}}),
$ which is the change in the distance after and before the mode 
bypassing, as a function of the dephasing strength $\gamma$.
In the regime  $\gamma\sim J$, both population and coherence in the site-basis remain coupled. Due to the coherence contribution, there is an overall enhancement in the distance measure after the mode-bypassing. As $\gamma$ increases, the quantity $\Delta$ decreases due to the negligible contribution from coherence. For the plots, we use the parameters $L=11$, $J=1$,  and for (a) $\gamma=1$. 
}
\label{dephasing_weak}
\end{figure}

\begin{figure}
    \centering
\includegraphics[width=0.45\columnwidth]{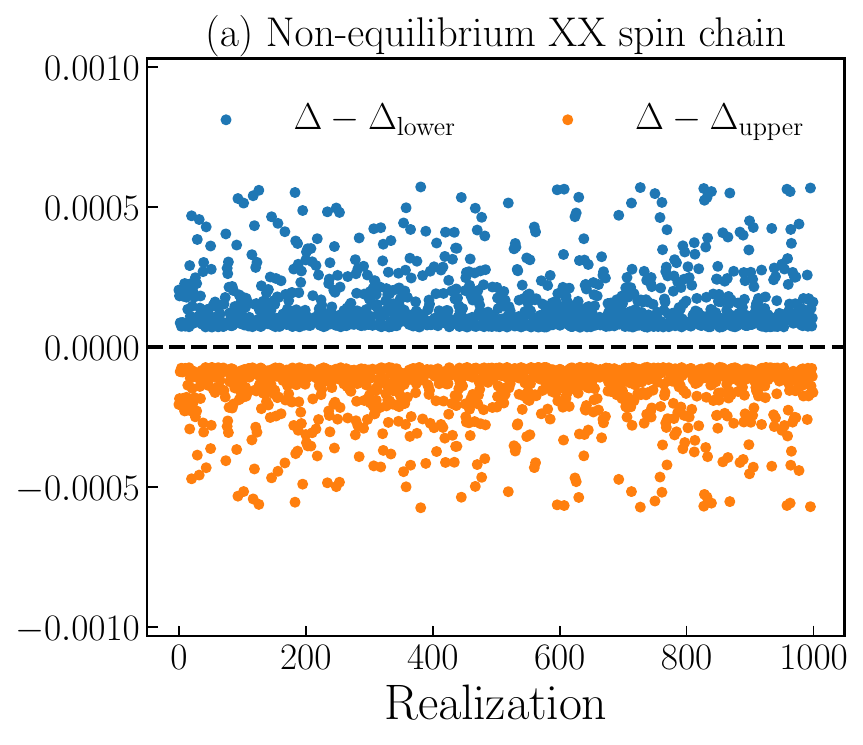} \includegraphics[width=0.435\columnwidth]{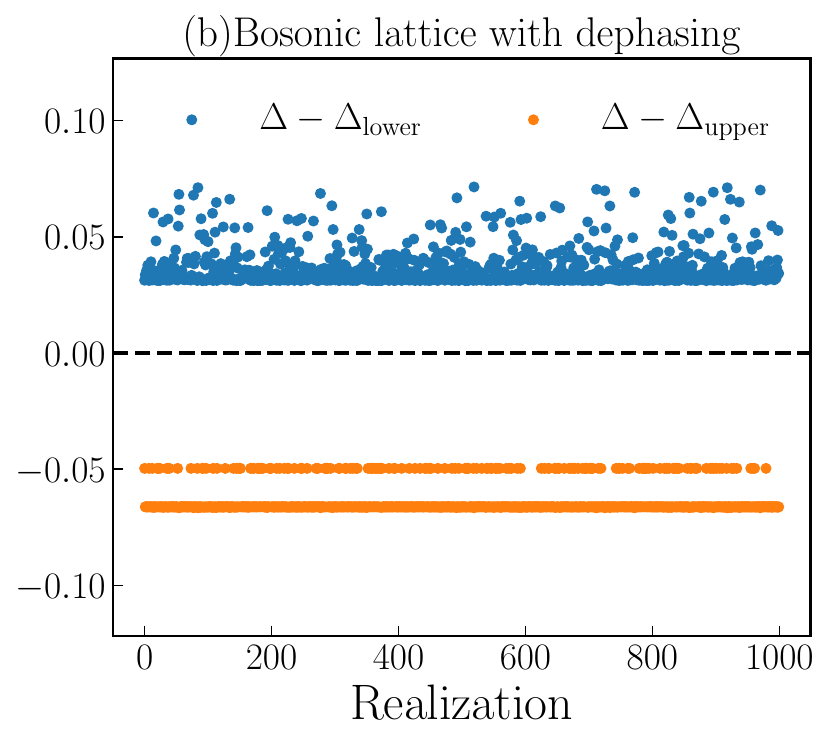}
    \caption{
Numerical verification of the bounds
(a) for the non-equilibrium XX spin chain model [Example 2 of the main text], and (b) for the bosonic lattice with dephasing [Example 3 of the main text]. For (a) we choose the initial operator $\mathcal{O}=
\sum_{\alpha_i\in\{x,y,z\}}
\bigotimes_{i=1}^{4}
\sigma_i^{\alpha_i},
$
and construct the transformed operator $\widetilde{\mathcal{O}}$ using the mode-bypassing procedure, where we bypass the SDM. We vary certain parameters of the setup randomly. For all such random realizations, as can be seen, the bounds are always satisfied. The numerical parameters are chosen as $g=0.05$, $\gamma_1=0.2$, $\gamma_N=0.5$, $\epsilon_{1,2,3,4}=(1,0.8,0.6,0.5)$, $T_1=1$, and $\mu_1=\mu_N=0$, while $T_N$ is varied randomly in the range $T_N\in[1.1,5]$. 
For (b), we take the population operator of the second site, $\mathcal{O}=n_2$, as the initial observable and construct the transformed operator by eliminating the SDM following the mode-bypassing procedure. The bounds are once again satisfied for all the random realizations. In this case, we use $J=1$ and vary the dephasing strength randomly in the range $\gamma\in[1,5]$.
}
\label{bound_numerical_check}
\end{figure}

\section{Numerical Verification of the Bounds in Eq.~(12)}
\label{AppendixF}
In the main text, we provided a lower and upper bound on the quantity $
\Delta=\mathcal{D}_{\rm dd}(\widetilde{\mathcal{O}},\mathcal{O}_{\rm ss})
-\mathcal{D}_{\rm dd}(\mathcal{O},\mathcal{O}_{\rm ss}),
$ which quantifies the change in the distance value after and before the mode-bypassing procedure. It satisfies the bounds
$\Delta_{\rm lower}\leq \Delta \leq \Delta_{\rm upper}$, 
where $\Delta_{\rm lower}=-\mathrm{Re}\,\big[\mathrm{Tr}(U_A^{\dagger}B)\big]$ and $\Delta_{\rm upper}=
-\mathrm{Re}\,\big[\mathrm{Tr}(U_{A-B}^{\dagger}B)\big]$, as given in Eq.~\eqref{inequality}. Recall that, 
$A=\mathcal{O}-\mathcal{O}_{\rm ss},
B=\mathrm{Tr}(l_1\mathcal{O})\,r_1,
$
with $r_1$ and $l_1$ denoting the right and left slow-mode eigenoperators, respectively. In this section, we demonstrate the validity of both the lower and the upper bounds for example 2 and example 3, considered in the main text.  In Fig.~\ref{bound_numerical_check} (a) and (b) we demonstrate the validity of both lower and upper bounds for the non-equilibrium XX spin chain [Example 2 of the main text] and for the bosonic lattice with dephasing [Example 3 of the main text]. We plot results for 1000 random realizations by randomly varying temperature $T_N$ for (a) and dephasing strength $\gamma$ for (b). In Fig.~\ref{bound_numerical_check} (a) and (b), we consider the initial operators
$\mathcal{O}=\sum_{\alpha_i\in\{x,y,z\}}\bigotimes_{i=1}^{4}\sigma_i^{\alpha_i},$
and
$\mathcal{O}=n_2$, respectively,  
both of which possess a nonzero overlap with the SDM of the Liouvillian $\mathcal{L}$. Using the mode-bypassing procedure, we construct the transformed operator $\widetilde{\mathcal{O}}$. To verify the bounds, we plot $\Delta-\Delta_{\rm lower}$ and $\Delta-\Delta_{\rm upper}$ for 1000 random realizations and observe the validity of Eq.~\eqref{inequality} in all cases.
 \end{document}